\documentclass [a4paper,12pt]{article}
\usepackage{amssymb}
\usepackage{graphicx}
\usepackage{amsmath}
\usepackage{fancybox}
\usepackage{color}
\usepackage{geometry}

\newcommand{\be}{\begin{equation}}
\newcommand{\ee}{\end{equation}}
\newcommand{\bea}{\begin{eqnarray}}
\newcommand{\eea}{\end{eqnarray}} 
\begin{document}
\setlength{\baselineskip}{18pt}
\begin{titlepage}

\begin{flushright}
NITEP 134         
\end{flushright} 
\vspace{1.0cm}
\begin{center}
{\LARGE\bf On the vacuum structure of gauge-Higgs unification models} 
\end{center}
\vspace{25mm}

\begin{center}
{\large

Yuki Adachi, C.S. Lim$^a$ and Nobuhito Maru$^{b, c}$
}
\end{center}

\vspace{1cm}

\centerline{{\it
Department of Sciences, Matsue College of Technology,
Matsue 690-8518, Japan.}}

\centerline{{\it
$^{a}$
Tokyo Woman's Christian University, Tokyo 167-8585, Japan }} 

\centerline{{\it 
$^b$
Department of Mathematics and Physics, Osaka Metropolitan University,
Osaka 558-8585, Japan}}

\centerline{{\it 
$^c$
Nambu Yoichiro Institute of Theoretical and Experimental Physics (NITEP), }} 
\centerline{{\it Osaka Metropolitan University,  
Osaka 558-8585, Japan}}
%
%

\vspace{2cm}
\centerline{\large\bf Abstract}
\vspace{0.5cm}

In this paper, we discuss the vacuum structure of the gauge-Higgs unification theory, which is one of the attractive candidates of physics beyond the standard model. This scenario has a remarkable feature, namely it has infinitely degenerate vacua due to the characteristic 
periodic potential of the Higgs filed, to be identified with the extra space component of the higher dimensional gauge field. We address a question, whether to form the superposition of such degenerate vacua, like the $\theta$-vacuum in QCD, is necessary or not, in order to realize the true vacuum state of the theory. We derive gauge field configuration which describes the transition between neighboring vacua, like the instanton (or anti-instanton) solution in QCD, and the corresponding Euclidean action in two models. In a simplified 2-dimensional U(1) model, the derived configuration to describe the transition is shown to have finite Euclidean action, and accordingly the ``$\theta$-vacuum" and the resultant ``$\theta$-term" are formulated. In a realistic 5D U(1) model, however, the gauge field configuration to describe the transition is shown to have infinite Euclidean action and therefore the tunneling probability between the degenerate vacua vanishes. Thus the superposition of the degenerate vacua is not necessary.

\end{titlepage}


\section{Introduction} 

In spite of its great success, the standard model of particle physics has some theoretical problems, which led the efforts to seek the theories beyond the standard model (BSM). One of such problems is well-known gauge hierarchy problem, i.e. the problem that in order to keep the weak scale $M_{W}$ small compared to the cutoff scale of the model under the quantum correction, some unnatural tremendous fine tuning of the theory parameter is necessary. This problem is recast into the one how to evade the quadratically divergent quantum correction to the Higgs mass-squared. 

In this paper we focus on the gauge-Higgs unification (GHU) scenario \cite{Manton, Hosotani}, where the Higgs field is identified with the extra-space component of some higher dimensional gauge field. To be precise, for instance in the simplest case of the gauge theory in the 5-dimensional (5D) space-time with gauge field $A_{M} = (A_{\mu}, A_y)$, the Kaluza-Klein (KK) zero-mode of $A_y$, $A^{(0)}_y$ behaves as the Higgs field. Since the mass term of gauge field never suffers from the divergent quantum correction in general, the gauge hierarchy problem is solved by virtue of the higher dimensional gauge symmetry \cite{HIL}. 

When the extra space is non-simply-connected space such as circle S$^{1}$ in the case of 5D gauge theory,  novel features emerge. 
If the extra space is not compactified, i.e. in the decompactification limit, the Higgs mass is never generated, not only at the classical level, but also at the quantum level, since the Higgs mass-squared term contradicts with the (higher dimensional) local gauge invariance. Interestingly, however, when the extra space is compactified on the S$^{1}$, there appears a potential for $A^{(0)}_y$, $V(A^{(0)}_{y})$, though the coefficient of each term is finite and the solution to the hierarchy problem is still valid \cite{HIL}. Furthermore, the potential is known to be a periodic function of $A^{(0)}_y$ with a period $1/gR$, as a specific feature of GHU. To be concrete, e.g. in the 5D QED model, assuming the electron is massless for the sake of simplicity \cite{HIL, Antoniadis, Kubo}, the potential is given by 
\be 
\label{1.1}
V(A^{(0)}_{y}) = \frac{3}{32\pi^{7}}\frac{1}{R^{5}}\sum_{w = 1}^{\infty} \ \frac{\cos (2\pi g A^{(0)}_{y}Rw)}{w^{5}}, 
\ee 
where the electric charge of the electron has been written as $g$ (instead of $e$ in the ordinary 4D QED). We easily confirm that the potential vanishes at the decompactification limit $R \to \infty$ as we expected. 

At the first glance, these novel features seem to contradict with the local gauge symmetry of the theory, since the potential, when regarded as a local operator, is not invariant under the local gauge transformation. (This is why the photon never gets its mass even under the quantum correction in the ordinary 4D QED.) What's really happening is that the potential $V(A^{(0)}_{y})$ can be understood as a function of a non-local operator, i.e. the Wilson-loop $W$ along the extra dimension S$^{1}$:  
\be 
\label{1.2} 
W = e^{i \int_{0}^{2\pi R} gA_{y} dy}  = e^{i 2\pi R gA^{(0)}_{y}},  
\ee 
where $g$ is a generic gauge coupling constant. The second equality comes from the fact that in the line integral of the KK mode expansion of $A_{y}$, $A_{y} = \sum_{n} A^{(n)}_{y}(t)e^{i \frac{n}{R}y}$, only the KK zero mode survives, since $\int_{0}^{2\pi R} e^{i \frac{n}{R}y} \ dy = 0$ for $n \neq 0$. As is well-known, the Wilson-loop $W$ is invariant under the local gauge transformation, $gA_{y} \to gA'_{y} = gA_{y} - \partial_{y}\lambda$ with gauge transformation parameter $\lambda$, which is periodic in $y$: 
\be 
\label{1.3} 
\lambda = \sum_{n} \lambda^{(n)}(t)e^{i \frac{n}{R}y}.   
\ee 
This gauge transformation clearly leaves the KK zero mode $A^{(0)}_{y}$ intact. In this way the potential $V(A^{(0)}_{y})$ is induced, being consistent with the gauge symmetry. Let us notice that, just as in the case of Aharonov-Bohm effect, the Wilson-loop is physically meaningful since the circle is non-simply-connected manifold and the line integral cannot be shrunk to a point. 

A remarkable thing in the GHU scenario is that the periodicity of the potential with the period $\frac{1}{gR}$ implies the presence of infinitely degenerate vacua, say $A^{(0)}_{y} = \nu \cdot \frac{1}{gR} \ (\nu: {\rm integer})$. Such characteristic vacuum structure of GHU models reminds us the rich vacuum structure of QCD and the resultant $\theta$-vacuum, which is obtained by the superposition of the infinitely degenerate vacua with suitable weight factors. The resultant ``$\theta$-term" implies an interesting physical consequence called strong CP problem. Thus, it will be natural to address a question whether a similar superposition of the infinitely degenerate vacua is necessary also in the GHU models of our interest, to realize the true vacuum. 

The transitions between different vacua are described by instanton (or anti-instanton) solutions in the case of QCD, whose Euclidean action determines the tunneling probability between those different vacua. Therefore, the central issues in this paper is what kind of field configuration describes the transition between different vacua and whether the corresponding tunneling probability is non-vanishing or not in the GHU models.   
If we get a non-vanishing probability, we clearly have to perform the superposition of the infinitely degenerate vacua as in the case of QCD to get a true vacuum of the model. Otherwise, such procedure will not be necessary and it will suffice to work in one of such infinitely degenerate vacua, the approach we usually take.

\section{An analysis in a simplified 2D U(1) model} 

The novel structure of the vacuum state, attributed to the periodic potential of the Higgs field in GHU scenario, already arises in a simplified Abelian U(1) gauge model. Thus, in this section we investigate the vacuum structure in the extremely simplified framework of U(1) GHU model in 2D space-time, with an extra-space alone as the spatial dimension, which is a circle S$^{1}$ of radius $R$. 

The 2D space-time is denoted by the coordinates $(x^{0}, y) \ (x^{0} = t)$ with $y$ being extra-space coordinate and the corresponding gauge fields are $(A_{0}, A_{y})$. The action $S$ is just the one for pure gauge theory including the potential $V(A^{(0)}_{y})$: 
\be 
\label{2.1} 
S = \int_{-\infty}^{\infty} dt \int_{0}^{2\pi R}dy \ \bigl\{-\frac{1}{2}F_{0y}F^{0y} - V(A^{(0)}_{y}) \bigr\}. 
\ee 
$V(A^{(0)}_{y})$ is expected to be induced radiatively, e.g., through 1-loop quantum effect due to the charged fermion, as in the 5D QED \cite{HIL}. Such matter field, however, is not included in this action, since it has nothing to do with the analysis of the vacuum structure of this theory given below. 

The radiatively induced effective potential in the 2D GHU model is given by \cite{Maru} 
\be 
\label{2.4}
V(A^{(0)}_{y}) = \frac{1}{2\pi^{3} R^{2}}\sum_{w = 1}^{\infty} \ \frac{\cos (2\pi g A^{(0)}_{y}Rw)}{w^{2}},   
\ee 
where $1/R^{5}$ and $1/w^{5}$ in the potential for the case of 5D QED, Eq.(\ref{1.1}), are replaced by $1/R^{2}$ and $1/w^{2}$, respectively because of the difference of the space-time dimension, together with a suitable change of the coefficient factor.

For the convenience in the discussions below, we make a translation, $A^{(0)}_{y} \to A^{(0)}_{y} +  \frac{1}{2gR}$, so that the potential is now written as 
\be 
\label{2.4a}
V(A^{(0)}_{y}) = \frac{1}{2\pi^{3} R^{2}}\sum_{w = 1}^{\infty} \ (-1)^{w} \frac{\cos (2\pi g A^{(0)}_{y}Rw)}{w^{2}}.  
\ee  
This potential is clearly periodic in $A^{(0)}_{y}$ with a period $\frac{1}{gR}$ and we have infinitely degenerate vacua at $A^{(0)}_{y} = \nu \cdot \frac{1}{gR} \ (\nu: {\rm integer})$. Thus, we address a question whether the superposition of the infinitely degenerate vacua like the $\theta$-vacuum in QCD is necessary in this GHU model.    

In the case of QCD, the degenerate vacua are denoted by a winding number, which specifies the homotopy class $\pi_{3}(SU(2))$ of the mapping ${\rm S}^{3} \to$ SU(2). Similarly, in our model infinitely degenerate vacua are denoted by a winding number $\nu$ given by a line integral, which is nothing but the Wilson-line phase,   
\be 
\label{2.5}
\nu = \frac{1}{2\pi} \int_{0}^{2\pi R} gA_{y} \ dy. 
\ee 
The integer $\nu$ specifies the homotopy class $\pi_{1}(U(1))$ of the mapping ${\rm S}^{1} \to$ U(1), where S$^{1}$ is the extra space. In fact, the $\nu$-th vacuum is given by a pure gauge configuration 
\be 
\label{2.6}
gA^{(0)}_{y} = U^{\dagger}i\partial_{y}U \ \ (U = e^{-i\frac{\nu}{R}y}),  
\ee
where $U = e^{-i\frac{\nu}{R}y}$ determines the mapping ${\rm S}^{1} \to$ U(1), with $\nu$ standing for how many times the unit circle on the complex plane representing U(1) is wrapped when $y$ goes from 0 to $2\pi R$.  

Our task is to find out the field configuration in this model corresponding to the instanton (or anti-instanton) solution in QCD, which is the classical solution to the Euclidean equation of motion and provides the minimum Euclidean action for a given homotopy class. To be more precise, we attempt to find the classical solution which describes the transition from the vacuum with the winding number $\nu = 0$ at $t \to - \infty$ to another vacuum with, say $\nu = 1$ at $t \to \infty$.  
To make the analysis simple, relying on the gauge invariance of the action, we first move to the gauge where $A_{y} = 0$ except for its KK zero mode $A^{(0)}_{y}$. This procedure is realized by a gauge transformation, $gA_{y} \to gA'_{y} = gA_{y} - \partial_{y}\lambda$ (see Eq.(\ref{1.3})), by choosing  
\be 
\label{2.7}
\lambda^{(n)} = -i \frac{R}{n}gA^{(n)}_{y} \ \ (n \neq 0).
\ee 
Let us note that the KK zero mode $A^{(0)}_{y}$ cannot be ``gauged away". Accordingly, $A_{0}$ also undergoes a gauge transformation $A_{0} \to A'_{0}$. 
Now the mixing between $A'_{0}$ and $A'_{y}$ through $F'_{0y}$ disappears since the non-zero KK modes of $A_{y}$ have been eliminated and zero modes $A^{(0)}_0$ and $A^{(0)}_y$ do not mix from the beginning as $\partial_{y}A^{(0)}_{0} = 0$. Hence, rewriting $A'_{0}, \ A'_{y}$ simply as $A_{0}$ and $A_{y} = A^{(0)}_{y}(t)$, the Euclidean action corresponding to the original action Eq.(\ref{2.1}) reduces to 
\be 
\label{2.8} 
S_{E} = \int_{-\infty}^{\infty} dt \int_{0}^{2\pi R}dy \ \frac{1}{2}\left( \frac{\partial A_{0}}{\partial y} \right)^{2}  
+ (2\pi R) \int_{-\infty}^{\infty} \ \bigl\{ \frac{1}{2}\left( \frac{dA^{(0)}_{y}}{dt}\right)^{2} + V(A^{(0)}_{y}) \bigr\} \ dt. 
\ee 
The (Euclidean) equation of motion for $A^{(0)}_{y}$ derived from this action is equivalent to that of the 1-dimensional motion of a particle with an unit mass under a potential 
\be 
\label{2.9} 
- V(A^{(0)}_{y}) \simeq  \frac{1}{2\pi^{3}R^{2}} \ \left[ \cos (2\pi g A^{(0)}_{y}R) - 1 \right], 
\ee 
where the potential is simplified by taking only the major contribution from the sector of $w = 1$ in Eq.(\ref{2.4}) and some suitable constant has been added so that $V(0) = 0$, which does not affect the essence of our argument. (If necessary, we can invoke numerical calculation by using the exact formula for the potential.) 

The transition from the vacuum $\nu = 0$ at $t \to - \infty$ to the vacuum $\nu = 1$ at $t \to \infty$ is described by a solution to the equation of motion with initial conditions $A^{(0)}_{y} = 0$ and $\frac{dA^{(0)}_{y}}{dt} = 0$ at $t = - \infty$. Therefore the conserved Euclidean energy (density) $\epsilon$ should vanish 
\be 
\label{2.10}
\epsilon = \frac{1}{2}\left( \frac{dA^{(0)}_{y}}{dt} \right)^{2} - V(A^{(0)}_{y}) = 0, 
\ee 
since $- V(A^{(0)}_{y}) = 0$ for $A^{(0)}_{y} = 0, \ \frac{1}{gR}$. Thus, $\frac{dA^{(0)}_{y}}{dt} = \sqrt{2V(A^{(0)}_{y})}$, and the classical solution to the equation of motion is obtained as follows:  
\be 
\label{2.11} 
\int_{\frac{1}{2gR}}^{A^{(0)}_{y}} \ \frac{d\tilde{A}^{(0)}_{y}}{\sqrt{\frac{1}{\pi^{3}R^{2}} \ \left[1 - \cos (2\pi g \tilde{A}^{(0)}_{y} R) \right] }} \simeq t \ \to \ A^{(0)}_{y} \simeq \frac{2}{\pi gR} \tan^{-1} e^{\sqrt{\frac{2}{\pi}} gt}.  
\ee 
The solution Eq.(\ref{2.11}) satisfies $A^{(0)}_{y}(-\infty) = 0, \ A^{(0)}_{y}(\infty) = \frac{1}{gR}$ (by choosing the values of the multi-valued function as $\tan^{-1} 0 = 0, \ \tan^{-1}\infty = \frac{\pi}{2}$), and therefore is confirmed to describe the transition between the neighboring vacua $\nu = 0 \to \nu = 1$. (By choosing other values for the multi-valued function, we can get the solutions to describe the transitions between the vacua, whose winding numbers differ not by 1 unit.)

The part of the action concerning $A^{(0)}_{y}$ given by this solution is easily found by use of Eq.(\ref{2.10}) (without using the explicit form of the solution Eq.(\ref{2.11})) to be 
\bea 
\label{2.12} 
S_{E} &=& (2\pi R) \int_{-\infty}^{\infty} \ \bigl\{ \frac{1}{2}\left(\frac{dA^{(0)}_{y}}{dt}\right)^{2} + V(A^{(0)}_{y}) \bigr\} \ dt \nonumber \\ 
&=& (2\pi R) \int_{-\infty}^{\infty} \ \left( \frac{dA^{(0)}_{y}}{dt}\right)^{2} \ dt = (2\pi R) \int_{-\infty}^{\infty} \ \sqrt{2V(A^{(0)}_{y})} \frac{dA^{(0)}_{y}}{dt} \ dt \nonumber \\ 
&\simeq & (2\pi R) \int_{0}^{\frac{1}{gR}} \ \sqrt{\frac{1}{\pi^{3}R^{2}} \ \left[1 - \cos (2\pi g A^{(0)}_{y}R) \right] } \ dA^{(0)}_{y}  \nonumber \\ 
&=&  \frac{4\sqrt{2} \pi^{-\frac{3}{2}}}{gR}.  
\eea  
On the other hand, the solution to the equation of motion for $A_{0}$ is clearly $A_{0} = 0$, as its initial conditions are  $A_{0} = \frac{\partial A_{0}}{\partial t} = 0$, and therefore the part of the action concerning $A_{0}$ also vanishes. Thus (\ref{2.12}) is actually the Euclidean action for the whole system, obtained using the solution to the equation of motion. As the conclusion, the semi-classical (WKB) approximation of the tunneling amplitude from the sector of $\nu = 0$ to that of $\nu = 1$ turns out to be non-vanishing:  
\be 
\label{2.12a} 
e^{- S_{E}} \simeq e^{- \frac{4\sqrt{2} \pi^{-\frac{3}{2}}}{gR}}. 
\ee

Since the Euclidean action is finite, we should form ``$\theta$-vacuum" as the real vacuum state mimicking the one in QCD: 
\be 
\label{2.13} 
| \theta \rangle = \sum_{\nu} e^{-i \nu \theta} | \nu \rangle, 
\ee 
where $| \nu \rangle$ stands for the vacuum state with winding number $\nu$.   

As the solution Eq.(\ref{2.11}) connects the vacua with $\nu$'s, different by one unit, it itself has a winding number 1, which is denoted by the line integral of the gauge field along the boundary of the 2D space-time (accompanied by a factor $g/(2\pi)$), as is easily seen by noting $A^{(0)}_{y}(t = \infty) - A^{(0)}_{y}(t = - \infty) = 1/(gR)$ (see Eq.(\ref{2.11})) and $A_{0} = 0$. This argument can be generalized to the solution with arbitrary winding number, and by use of the Stokes' theorem, the winding number $\nu$ is given by a gauge invariant 2D topological term (corresponding to the $\int d^{4}x \ F_{\mu \nu}\tilde{F}^{\mu \nu}$ in 4D space-time):  
\be 
\label{2.14} 
\nu = \frac{g}{2\pi} \int_{-\infty}^{\infty} dt \int_{0}^{2\pi R}dy \ F_{0y}.   
\ee

Hence, by use of Eq.(\ref{2.14}), we easily see that the effect of adopting the theta vacuum is equivalent to the introduction of the following ``$\theta$-term" in the lagrangian: 
\be 
\label{2.15}
{\cal L}_{\theta} = \frac{g}{2\pi} \theta \int_{-\infty}^{\infty} dt \int_{0}^{2\pi R}dy \ F_{0y}. 
\ee 

Interestingly, if the space coordinate $y$ is replaced by $x_{1} = x$ and the potential term is ignored, this system is equivalent to the well-known Schwinger's model \cite{Schwinger}, 2D QED, which has specific and remarkable features, such as the ``chiral" anomaly, given by exactly the same expression as the r.h.s. of (\ref{2.14}) and (\ref{2.15}), $\partial_{\mu}j_{5}^{\mu} = \frac{g}{2\pi} \epsilon^{\mu \nu}F_{\mu \nu} \ (\mu, \nu = 0,1)$. Let us note that the potential for the gauge field is meaningless for de-compactified space.

\section{An analysis in a realistic 5D U(1) model} 

Now we move to the discussion on the vacuum structure of a realistic model, 5D U(1) GHU model. 
In particular, we focus on the question whether the tunneling probability between neighboring vacua with winding numbers different by 1 unit stays finite as in the 2D GHU model or not. 

The 5D space-time is denoted by the coordinates $(x^{\mu}, y) \ (\mu = 0,1,2,3)$, with the extra-space being assumed to be a circle S$^{1}$ of radius $R$. The corresponding gauge fields are $(A_{\mu}, A_{y})$. The 5D action for the gauge fields is 
\be 
\label{3.1}
S = \int d^{4}x \int_{0}^{2\pi R}dy \ \bigl\{-\frac{1}{4}F_{\mu \nu}F^{\mu \nu} - \frac{1}{2}F_{\mu y}F^{\mu y} - V(A^{(0)}_{y}) \bigr\}, 
\ee 
where the radiatively induced potential $V(A^{(0)}_{y})$ due to the exchange of U(1)-charged fermion (with vanishing bulk mass) is given by 
Eq.(\ref{1.1}). 

As in the previous section, we make a translation, $A^{(0)}_{y} \to A^{(0)}_{y} +  \frac{1}{2gR}$, so that the potential is written as 
\be 
\label{3.2a}
V(A^{(0)}_{y}) = \frac{3}{32\pi^{7}}\frac{1}{R^{5}}\sum_{w = 1}^{\infty} \ (-1)^{w} \frac{\cos (2\pi g A^{(0)}_{y}Rw)}{w^{5}}.  
\ee 
It is simplified by taking only the dominant contribution of the sector $w = 1$ and adding a suitable constant so that $V(0) = 0$: 
\be 
\label{3.2b}
V(A^{(0)}_{y}) \simeq \frac{3}{32\pi^{7}}\frac{1}{R^{5}} \ \left( 1 - \cos (2\pi g A^{(0)}_{y}R) \right).  
\ee 

Again we adopt a gauge where $A_{y} = 0$ except for its KK zero-mode $A^{(0)}_{y}$, which cannot be gauged away. Then, as in the case of the 2D model discussed in the previous section, the mixing between $A_{\mu}$ and $A_{y}$ disappears and 
the Euclidean action in this choice of gauge reduces to 
\bea 
S_{E} &=& \int d^{4}x \int_{0}^{2\pi R}dy \ \bigl\{ \frac{1}{2} (F_{0i})^{2} + \frac{1}{4}(F_{ij})^{2} + \frac{1}{2}\left( \frac{\partial A_{\mu}}{\partial y} \right)^{2} \bigr\} \nonumber \\ 
&+& (2\pi R) \int d^{4}x \ \bigl\{ \frac{1}{2}\left( \frac{\partial A^{(0)}_{y}}{\partial t}\right)^{2} + \frac{1}{2}\left( \frac{\partial A^{(0)}_{y}}{\partial x^{i}}\right)^{2} + V(A^{(0)}_{y}) \bigr\},  
\label{3.3} 
\eea  
where $i, j = 1, 2, 3$ are indices corresponding to the 3D space. 

As in the case of 2D U(1) GHU model, this theory has infinitely degenerate vacua denoted by $A^{(0)}_{y} = \nu \cdot \frac{1}{gR} \ (\nu: {\rm integer})$. Our concern is the tunneling probability between these vacua. Especially, we focus on the transition from the vacuum with the winding number $\nu = 0$ at $t \to - \infty$ to the neighboring vacuum $\nu = 1$ at $t \to \infty$. Such transition is described by a solution to the Euclidean equation of motion with initial conditions $A^{(0)}_{y} = 0$ and $\frac{\partial A^{(0)}_{y}}{\partial t} = 0$ at $t = - \infty$. Since the initial conditions do not depend of $x^{i}$, the solution to the Euclidean equation of motion, 
\be 
\label{3.4} 
\frac{\partial^{2} A^{(0)}_{y}}{\partial t^{2}} + \frac{\partial^{2} A^{(0)}_{y}}{\partial x^{i2}} - \frac{d V(A^{(0)}_{y})}{d A^{(0)}_{y}} = 0, 
\ee 
should depend only on $t$ and can be written in a form $A^{(0)}_{y} = f(t)$. Plugging this into Eq.(\ref{3.4}), $f(t)$ is known to satisfy an equation, 
\be 
\label{3.6}
\frac{d^{2} f}{dt^{2}} - \frac{dV}{df} = 0, 
\ee 
under the initial conditions $f = \frac{df}{dt} = 0$ at $t = - \infty$. Just as in the case of 2D model, this mechanical system is equivalent to the system of a point particle with unit mass moving under the potential $- V$. 

As we did in the previous section, we impose a condition that the conserved Euclidean energy (density) $\epsilon$ vanishes:   
\be 
\label{3.7}
\epsilon = \frac{1}{2}\left( \frac{df}{dt} \right)^{2} - V(f) = 0,  
\ee 
which leads to $\frac{df}{dt} = \sqrt{2V(f)}$, and the solution to the equation of motion, satisfying $f = \frac{df}{dt} = 0$ at $t = - \infty$, is easily obtained as follows:  
\be 
\label{3.8} 
\int_{\frac{1}{2gR}}^{f} \ \frac{d\tilde{f}}{\sqrt{\frac{3}{16\pi^{7}}\frac{1}{R^{5}} \ \left[1 - \cos (2\pi g \tilde{f} R) \right] }} \simeq t \ \to \ A^{(0)}_{y} = f(t) \simeq \frac{2}{\pi gR} \tan^{-1} e^{\sqrt{\frac{3}{8\pi^{5}R^{3}}} gt}.   
\ee 
Since $A^{(0)}_{y}(-\infty) = 0$ and $A^{(0)}_{y}(\infty) = \frac{1}{gR}$, this solution describes the transition $\nu = 0 \to \nu = 1$.

The part of the action concerning $A^{(0)}_{y}$ given by this solution is easily found by use of Eq.(\ref{3.7}) to be 
\bea 
\label{3.9} 
S_{E} &=& (2\pi R)V_{3D} \int_{-\infty}^{\infty} \ \bigl\{ \frac{1}{2}\left(\frac{df}{dt}\right)^{2} + V(f) \bigr\} \ dt \nonumber \\ 
&=& (2\pi R) V_{3D} \int_{-\infty}^{\infty} \ \left( \frac{df}{dt}\right)^{2} \ dt \nonumber \\ 
&\simeq & (2\pi R) V_{3D} \int_{0}^{\frac{1}{gR}} \ \sqrt{\frac{3}{16\pi^{7}}\frac{1}{R^{5}} \ \left[1 - \cos (2\pi g fR) \right] } \ df  \nonumber \\ 
&=& \sqrt{\frac{6}{\pi^{7}R^{5}}}\frac{1}{g}V_{3D} = \frac{\sqrt{3}}{\pi^{4}}\frac{1}{g_{4}} \frac{V_{3D}}{R^{3}},   
\eea  
where $V_{3D}$ denotes the volume of the 3D space and $g_{4} = \frac{g}{\sqrt{2\pi R}}$ is the 4D (dimension-less) gauge coupling constant. 

On the other hand, the solution to the equation of motion for $A_{\mu}$ is clearly $A_{\mu} = 0$, since its initial conditions are  $A_{\mu} = \frac{\partial A_{\mu}}{\partial t} = 0$, and the part of the action concerning $A_{\mu}$ also vanishes. Thus Eq.(\ref{3.9}) is exactly  the Euclidean action for the whole system due to the solution to the equation of motion. 
Thus, in the limit $V_{3D} \to \infty$, $S_{E}$ also gets infinity, and therefore the tunneling amplitude between the neighboring vacua vanishes: 
\be 
\label{3.10}
\lim_{V_{3D} \to \infty} e^{- S_{E}} = 0. 
\ee 
Thus, the procedure we usually take, i.e. to assume that the vacuum state is one of the degenerate vacua with, say $\nu = 0$, is justified.

\section{Summary and discussion} 

In this paper, we discussed the novel feature of the GHU scenario, one of the attractive 
candidates of physics BSM, namely the infinitely degenerate vacua due to the characteristic 
periodic potential of the Higgs filed, to be identified with the extra space component of the higher dimensional gauge field. 
We discussed that the periodicity is the inevitable consequence of the fact that the Higgs field is interpreted to be a Wilson-loop phase along the extra dimension, which has a physical meaning when the extra space is non-simply-connected manifold like a circle S$^1$. 

We addressed a question, whether to form the superposition of such degenerate vacua, like the ``$\theta$-vacuum" in QCD, is necessary or 
not, in order to realize the true vacuum state of the theory. Especially, we derived gauge field configuration which describes the transition between neighboring vacua, like instanton (or anti-instanton) solution in QCD, and the corresponding Euclidean action, which determines the tunneling probability between the neighboring vacua.   

First, as the extremely simplified model, we discussed 2D U(1) GHU model. Each of the degenerate vacua is denoted by a topological winding number, corresponding to the homotopy class $\pi_{1}(U(1))$. We derived a configuration to describe the transition between neighboring vacua, and have shown that it has finite Euclidean action and therefore non-vanishing tunneling probability. Accordingly, the $\theta$-vacuum was formulated and the resultant ``$\theta$-term" was presented.  

Next, we discussed a realistic 5D U(1) GHU model. In contrast to the case of the 2D U(1) GHU model, though we could find the gauge field configuration to describe the transition between neighboring vacua, the corresponding Euclidean action turns out to be infinite. This means 
the superposition of the degenerate vacua is not necessary, and we can take one of the degenerate vacua as our vacuum state, the attitude we 
usually take.  

Though we have concentrated on the U(1) GHU models, the periodic Higgs potential is the common feature of GHU, shared also by non-Abelian GHU models, as long as the extra space is non-simply-connected manifold, such as S$^{1}$. The winding number $\nu$ in order to characterize the degenerate vacua is still given by the Wilson-loop phase. For instance, in the case of the minimal SU(3) electro-weak unified GHU model \cite{Kubo}, $\nu$ is written in terms of the line-integral of $A^{6}_{y}$ associated with the Gell-Mann matrix $\lambda_{6}$, which develops the 
VEV. Note that in this case the homotopy class $\pi_{1}$(U(1)) is for the mapping from S$^{1}$ to the U(1) generated by $\lambda_{6}$, the subgroup of SU(3) (If necessary, $\lambda_{6}$ can be diagonalized by a suitable gauge transformation).

\subsection*{Acknowledgments}

This work was supported in part by Japan Society for the Promotion of Science, Grants-in-Aid for Scientific Research, No.~15K05062.


\providecommand{\href}[2]{#2}\begingroup\raggedright\endgroup

\end{document}